\documentclass[11pt]{article}
\usepackage{moriond,epsfig}

\def\be{\begin{equation}}
\def\ee{\end{equation}}
\def\bea{\begin{eqnarray}}
\def\eea{\end{eqnarray}}

\begin{document}
\vspace*{2cm}
\title{MASS OF CLUSTERS IN SIMULATIONS}

\author{ A.V. MACCI\`O$^{1,2}$, G. MURANTE$^3$, S.A. BONOMETTO$^{1,2}$}
\address{$^1$ Dep. of Physics G. Occhialini, Milano Bicocca Univ.,
Piazza della Scienza 3, 20126 Milano (Italy) \\
$^2$ I.N.F.N. -- Sezione di Milano \\
$^3$ I.N.A.F., Osservatorio Astronomico di Torino - Torino (Italy)}

\maketitle\abstracts{We show that dark matter haloes, in n--body
simulations, have a 
{\it boundary layer} (BL) with precise features. In particular,
it encloses all dynamically stable mass while, outside it, dynamical 
stability is lost soon. Particles can pass through such BL, which 
however acts as a confinement barrier for dynamical properties. 
BL is set by evaluating kinetic and potential energies ($T(r)$ 
and $W(r)$) and calculating ${\cal R}=-2T/W$. Then, on BL, 
${\cal R}$ has a minimum which closely approaches a maximum of $w=
-d\log W/d\log r$. Such  $Rw$ ``requirement'' is consistent
with virial equilibrium, but implies further regularities.
We test the presence of a BL around haloes in spatially flat 
CDM simulations, with or without cosmological constant. We find that
the mass $M_c$, enclosed within the radius $r_c$, where the {\it Rw} 
requirement is fulfilled, closely  approaches the 
mass $M_{dyn}$, evaluated from the velocities of all particles within
$r_c$, according to the virial theorem. 
Using $r_c$ we can then determine
an individual density contrast $\Delta_c$ for each virialized halo, 
which can be compared with the "virial" density contrast $\Delta_v 
\simeq 178 \, \Omega_m^{\, 0.45}$ ($\Omega_m$: matter density
parameter) obtained assuming a spherically symmetric 
and unperturbed fluctuation growth. 
The spread in $\Delta_c$ is wide, and 
cannot be neglected when global physical quantities related to 
the clusters are calculated, while the average $\Delta_c$ is
$\sim 25\, \%$ smaller than the corresponding $\Delta_v$;
moreover if $M_{dyn}$ is defined from the radius linked to $\Delta_v$,
we have a much worse fit with particle mass then starting from {\it
Rw} requirement.
}

\section{Introduction}

Finding the volume where cluster materials are in virial equilibrium
might seem a tough and somewhat ambiguous task, as clusters are not
surrounded by vacuum and shall be however confined by unbound
background materials. Here we show that, in general, the transition
from cluster and external materials is well defined and takes
place just outside a {\it boundary layer} (BL), which, by itself,
is in virial equilibrium with no pressure pushes, but 
above which virial equilibrium is lost soon.
Particles may pass through such BL, in both directions,
but its overall virial balance keeps constant in time.
We find such BL assuming a spherical symmetry.
At a distance where the cluster ends up into unbound
materials, such requirement might seem excessive. Rather surprisingly,
however, we shall see that this is not a prohibitive restriction.

Previous work on this point was based on the assumption of
isolated fluctuation growth, starting from and preserving spherical symmetry
(Gott \& Rees 1975, Lahav et. al. 1991, Eke et al. 1996, 1998,
Brian \& Norman 1998). Results obtained in this way were then
applied to simulation analysis. On the contrary, in our work, the 
assumption of isolated growth is absent, while spherical symmetry is
tested, {\it a posteriori}, on simulated clusters. In this way
we find that actual clusters closely approach properties
predicted under such assumption.

\section{How to find {\it BL}}

Let us define {\it integral} potential and kinetic energy and 
their {\it virial ratio}:
$$
2T(<r) = {\sum_i}_{(r_i<r)} mv_i^2 ~,~~~~~~~~~~~
W(<r) = -{\sum_{i<j}}_{(r_{i,j}<r)} {Gm^2 \over r_{ij}} ~,~~~~~~~~~~
{\cal R}(<r) = -{2T(<r) \over W(<r)} ~.
\eqno (2.1)
$$
Similar quantities can be defined for $r$--intervals.
The $r$ dependence of $W$ can be outlined by performing the sums
in eq.~(2.1) starting from outside:
$$
W(<r) = -{\sum_{i}}_{(r_i<r)} {\sum_j}_{(r_j<r_i)} {Gm^2 \over r_{ij}}
= -{\sum_{i}}_{(r_i<r)} {\cal Z}(r_i)  .
\eqno (2.2)
$$
Here the last term is the very definition of $\cal Z$. 
Quite in general, a volume integral of $\rho(r)$ increases with $r$. 
Henceforth, 
${\cal Z}(r) = C (r/{\bar r})^{-w},
$
where C is a normalization constant evaluated 
at an arbitrary position $\bar r$ and $w (< 1)$
shall however depend on $r$. Accordingly, it must be
$$
r{d{\cal Z} \over dr} (r)= 
-\big[w + rw'\ln\big({r \over{\bar r}}\big)\big]{\cal Z}(r)  ~.
\eqno (2.4)
$$
Let us now suppose that there exists an interval $\Delta r =r_+ - r_-$,
(i) in virial equilibrium, (ii) subject to no pressure pushes inwards 
or outwards, (iii) inside which $w$ is constant. 
Owing to the conditions (i) and (ii), in this layer
$$
{\sum_i}_{(r_i \in \Delta r)}  mv^2_i -
 r_i {d{\cal Z} \over dr}(r_i) = 0~.
\eqno (2.5)
$$
Taking then into account the extra condition (iii), i.e.
that $w$ is constant, we can use eq.~(2.4) with $w'=0$ and the 
expression for the {\it virial ratio} in $\Delta_r$ to obtain that
$$
{\cal R} = w ~~~~~~,~~~~~~ {d{\cal R} \over dr} = 0 ~
\eqno (2.6)
$$
all along the interval $\Delta r$. 
{\it Viceversa}, if the eq.s~(2.6) are 
both fulfilled in a layer of depth $\Delta r$, such layer is at
rest and in virial equilibrium.
We thus define {\it boundary layer (BL)} a region of depth
$\Delta r$ satisfying eq.s~(2.6).
It can also be shown that no further
materials can be in virial equilibrium if we required that $w$ is
maximum in $\Delta r$ and therefore $w' < 0$ for $r>r_+$.

\section{Application to numerical simulations}

We seek BL for clusters, in simulations. We use two main sets of 
simulations, performed with different codes. The former set comprises
large volume simulations (box side=360$h^{-1}$ Mpc), run
with a parallel AP3M code developed by Gardini et al. (1999).
They describe a tilted CDM (TCDM) model ($n=0.8$) and a $\Lambda$CDM model
($\Omega_m=0.35$, $\Omega_{\Lambda}=0.65$) and were
also used in Macci\`o et al. (2001; simulations A and B respectively).
This set of simulations is meant to provide a significant
statistics, although their mass (180$^3$ particles) and
force resolution ($\sim 40\, $kpc P.E.) is limited.
The second set of simulations aims to provide
high--resolution clusters, using the PM ART code (Kravtshov, Klypin 
\& Khokhlov, 1997) and/or the public parallel 
tree--code GADGET (Springel et. al 2001).
Here we resolve clusters with more than 300.000 
particles within a radius of 2 $h^{-1}$ Mpc, with a mass resolution 
of 1.2 $ \times 10^{10} \Omega_m h^{-1} M_\odot$. 
With these codes we performed simulations C and D of $\Lambda$CDM 
($\Omega_m=0.3$ ,  $\Omega_{\Lambda}=0.7$) 
and TCDM ($n=0.8$) models, obtaining
6 high resolution clusters for each model.

To set the position of BL,
in each dark matter halo, we first calculate the virial 
ratio $\cal R$ in successive layers; then, we select points
where $\cal R$ has a minimum and (nearly) intersects $w$ 
(so to fulfill eq.s~2.6).
Such coincidence is shown in Fig.~1 for the best and worst cases we 
found in the simulations C and D. Even in worst case,
the coincidence between a minimum of $\cal R$ and a maximum 
of $w$ is neat. In simulations A and B, $w$ is not
so well traced, as it is obtained from differentiation,
and is only used to select among minima of $\cal R$, when
we have several of them. The procedure however works
in 97$\, \%$ of cases, allowing to find a precise BL.

Once the sphere confining cluster materials is set, we can 
evaluate the density contrast $\Delta_c$ and the mass $M_c$ thereinside.
In Fig.s~2 and 3, points give $\Delta_c$ and $M_c$ for all clusters in 
simulations A and B, respectively. They show that the
spread of $\Delta_c$ values is fairly wide. By subdividing
the $M_c$ abscissa in intervals of constant logarithmic width,
we evaluate the average density contrast in each of them,
to seek systematic trends with mass.

Owing to the spread of $\Delta_c$ values, their averages
are still subject to a significant uncertainty, shown, at the 1--$\sigma$
level, in the plots. 
There seems to be no evidence of any peculiar trend of density
contrasts with mass apart, perhaps, a modest indication of
an increasing density contrast, at very high scales, in $\Lambda$CDM.
It is therefore licit to consider the overall average among
$\Delta_c$'s. Such average is indicated by the continuous horizontal
line and compared with the "virial" density contrast $\Delta_v$, as 
given by eq.~(1.1). In both cases, $\Delta_v$ (dotted line) is well 
inside the range of the density contrasts we found; the average
$\Delta_c$, however, in both cases, is smaller than $\Delta_v$
by $\sim 25\, \%$.

A critical result of our analysis is however shown in Fig.s~4 and 5
(for simulations A and B, respectively).
In simulations, cluster masses can be comfortably evaluated
by summing up particle masses and
we compared this mass extimate with
$ M_{dyn}=  {{ \langle v^2 \rangle_{\{v,c\}} \cdot r_{\{v,c\}}}  \over G}
$, evaluated averaging over the velocities of the particles within
$r_{v,c}$ ($r_v$ is the radius whereinside the density contrast is $\Delta_v$).
This comparison yields
a reasonable coincidence between $M_v$ and $M_{dyn}$ as is shown
in Fig.s~4 and 5 by the dashed histograms (which
confirm the slight excess of $M_{dyn}$ $vs$. $M_v$ already noticed
by previous authors). In the same plots we also report the much better
fit between $M_{dyn}$ and $M_c$, obtained on the basis of the
setting of the BL (continuous line).  
The average values of $M_{dyn}/M_c$ are
$\sim 0.97 \pm 0.03$ for both models.

\section{Conclusions}

Our results are based on the mathematical definition of a boundary 
layer (BL), bordering a region satisfying the virial theorem,
although its density contrast has different values in different haloes.
The BL is found in numerical simulation,
around each cluster, and it is shown to separate bound cluster 
materials from
the surrounding medium. Boundary layers can be found in simulations
with sufficient dynamical range, and have been confirmed by inspections
of higher resolution simulations. 
The spread of the density contrasts
of individual clusters has been found to be fairly wide and, in
average, $\Delta_c$ are smaller by $\sim 25\, \%$ when compared to 
the so--called "virial" density contrast ($\Delta_v 
\simeq 178 \, \Omega_m^{\, 0.45}$).
No evidence of a systematic trend of $\Delta_c$ with $M_c$ was found.
The fit between dynamical masses and particle masses, obtained
with our technique, is however far better than the one obtained
starting from a fixed virial density contrast.

\section*{Acknowledgments}

We thank INAF for allowing us the CPU time 
to perform the ART simulation C and D at the CINECA consortium
(grant cnami44a on the SGI Origin 3800 machine).
GADGET simulations of high--resolution clusters have
been run on the 16 Linux PC Beowulf cluster at the Osservatorio
Astronomico di Torino.

\begin{figure}
\centerline{\mbox
{\epsfysize=6.0truecm\epsffile{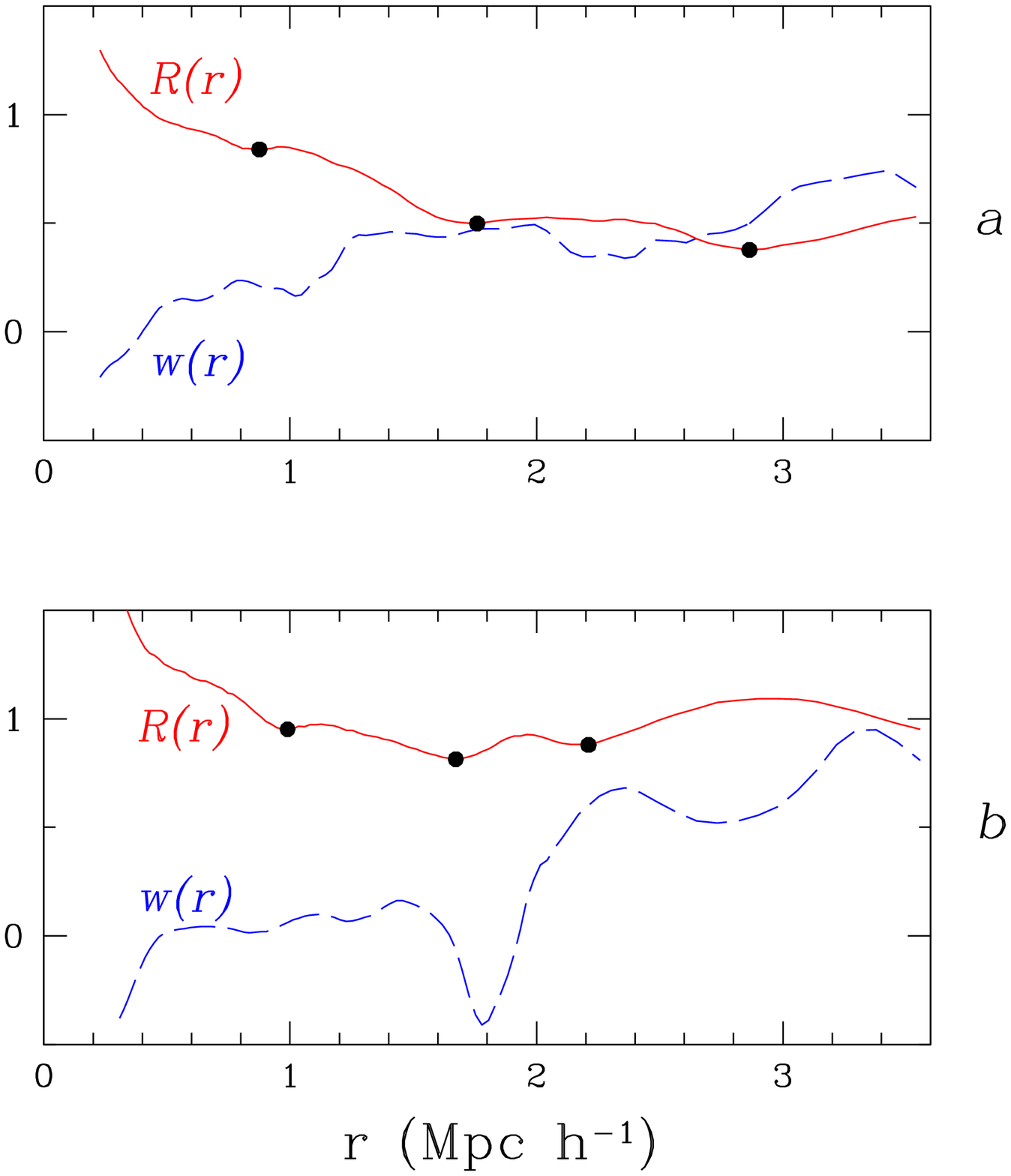}}
{\epsfysize=6.0truecm\epsffile{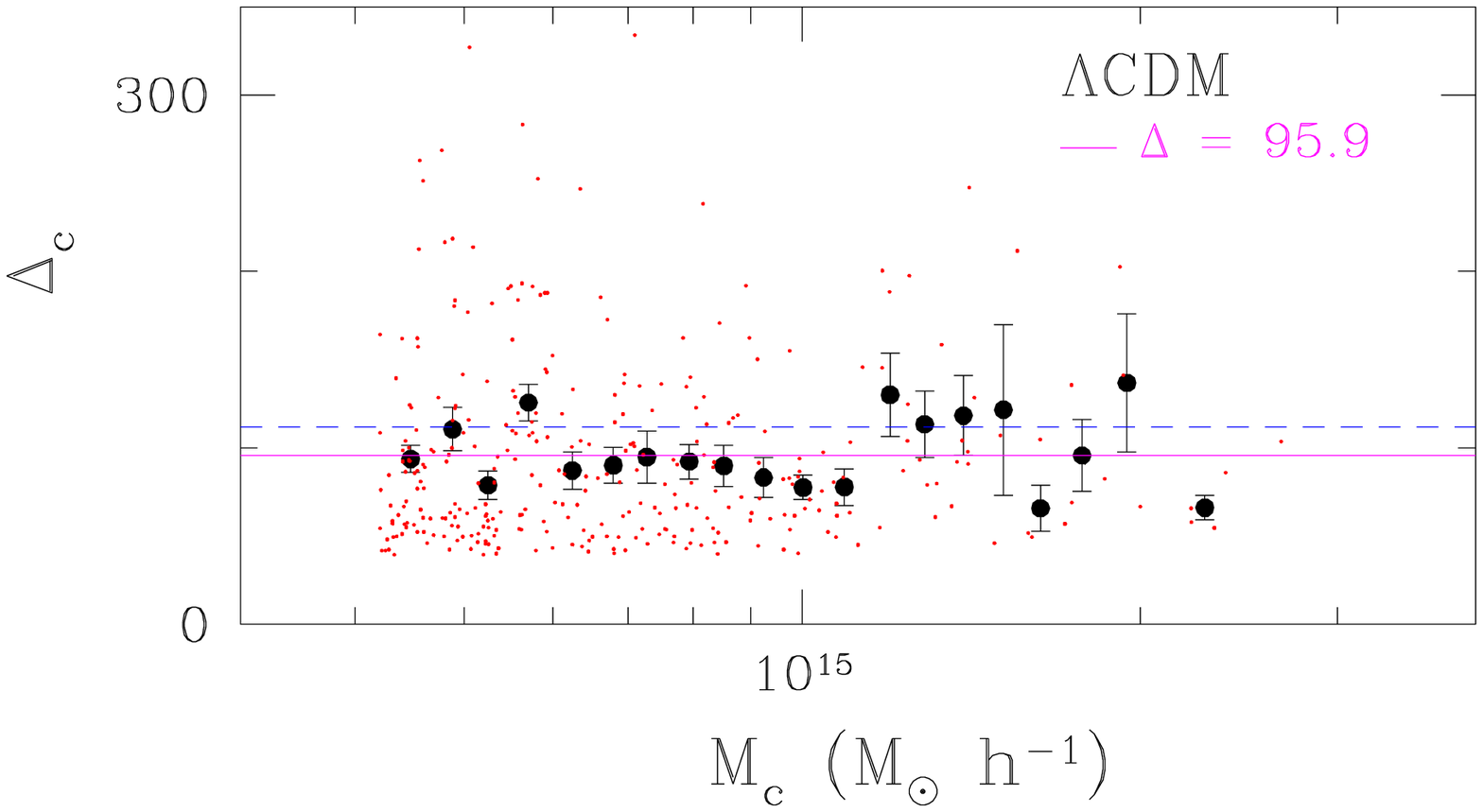}}
{\epsfysize=6.0truecm\epsffile{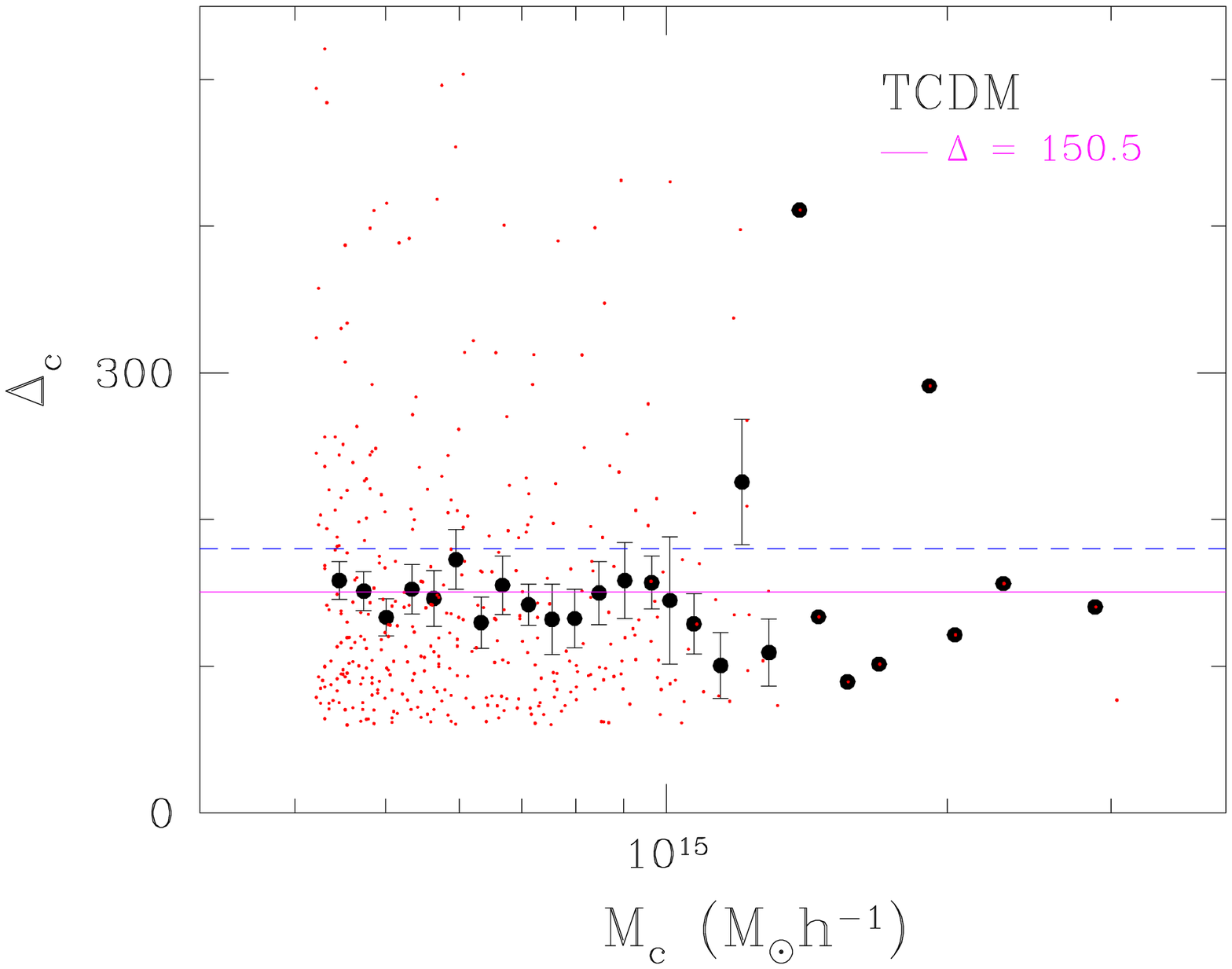}}}

\centerline{Fig. 1 ~~~~~~~~~~~~~~~~~~~~~~~~~~~~~~~~~~~~~~~~~~~~Fig.2 
~~~~~~~~~~~~~~~~~~~~~~~~~~~~~~~~~~~~~~~~~~~~Fig.3}
\end{figure}

\begin{figure}
\centerline{\mbox
{\epsfysize=5.0truecm\epsffile{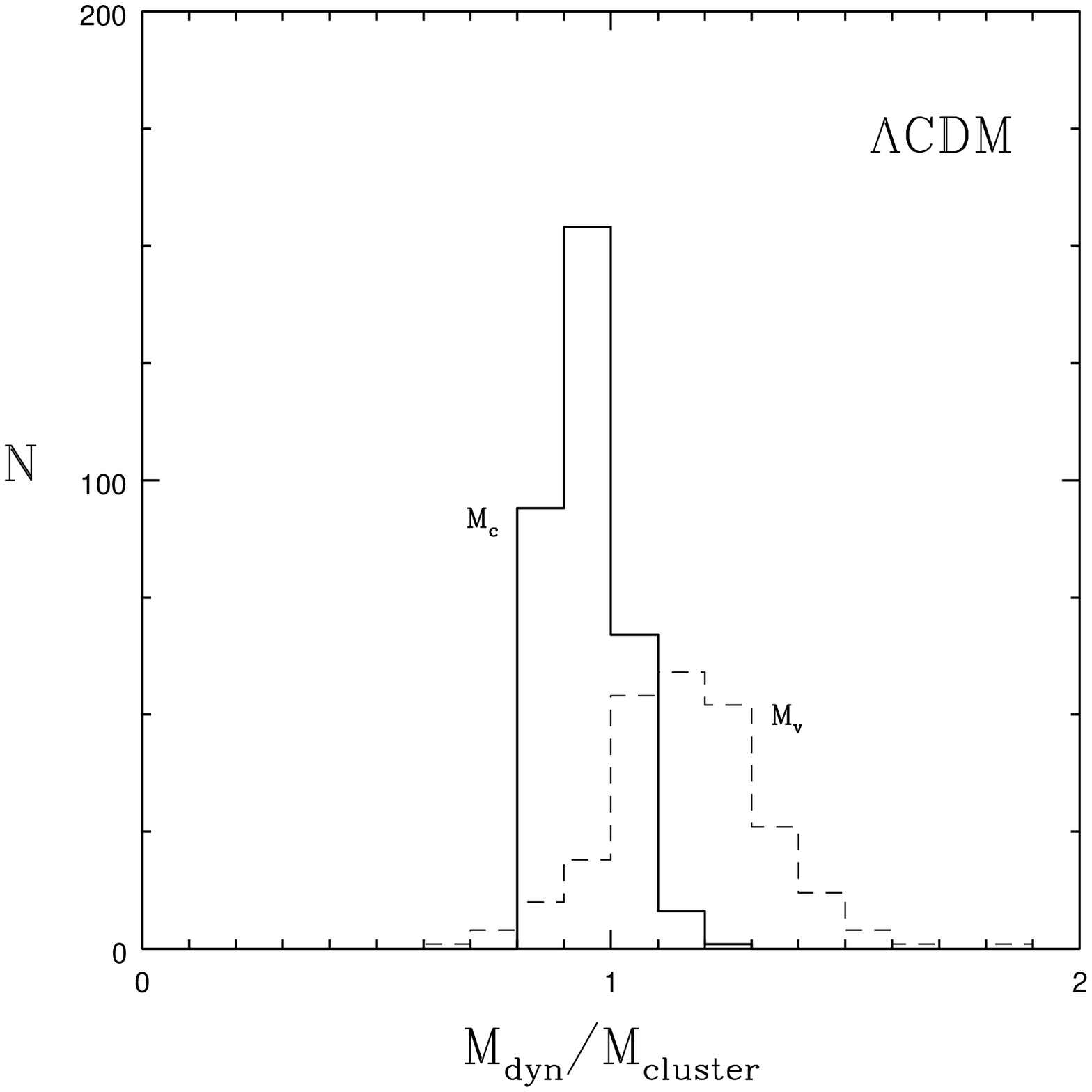}}
{\epsfysize=5.0truecm\epsffile{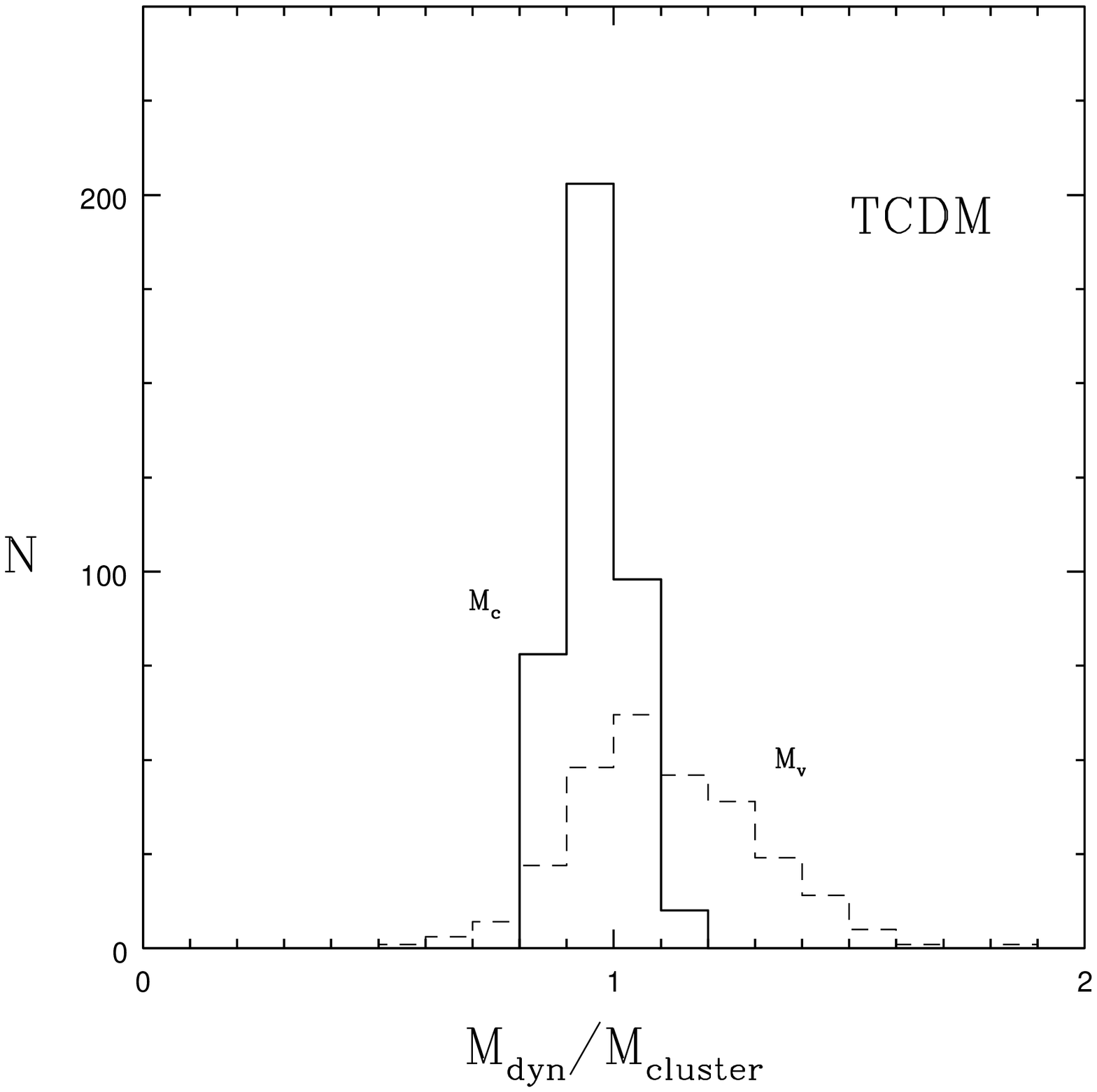}}
}

\centerline{Fig 4 ~~~~~~~~~~~~~~~~~~~~~~~~~~~~~~~~~~Fig.5}
\end{figure}

\end{document}